\documentclass[twocolumn]{aastex63}

\usepackage[ruled,noline]{algorithm2e}
\usepackage{algpseudocode}

\newcommand{\code}{\texttt}

\shorttitle{Non-Linear SNR Optimization}
\shortauthors{Thompson and Marois}
\graphicspath{{./}{figures/}}

\begin{document}

\title{Improved Contrast in Images of Exoplanets using Direct SNR Optimization}

\correspondingauthor{William Thompson}
\email{wthompson@uvic.ca}

\author[0000-0001-5684-4593]{William Thompson}
\affiliation{
    University of Victoria,
    Department of Physics and Astronomy, \\
    3800 Finnerty Rd,
    Victoria, BC V8P 5C2, Canada}
\affiliation{
    National Research Council of Canada Herzberg,\\
    5071 West Saanich Rd,
    Victoria, BC, V9E 2E7, Canada}

\author[0000-0002-4164-4182]{Christian Marois}
\affiliation{
    National Research Council of Canada Herzberg,\\
    5071 West Saanich Rd,
    Victoria, BC, V9E 2E7, Canada}
\affiliation{
    University of Victoria,
    Department of Physics and Astronomy, \\
    3800 Finnerty Rd,
    Victoria, BC V8P 5C2, Canada}



\begin{abstract}

    Direct imaging of exoplanets is usually limited by quasi-static speckles. These uncorrected aberrations in a star's point spread function (PSF) obscure faint companions and limit the sensitivity of high-contrast imaging instruments. Most current approaches to processing differential imaging sequences like angular differential imaging (ADI) and spectral differential imaging (SDI) produce a self-calibrating dataset that are combined in a linear least squares solution to minimize the noise. Due to temporal and chromatic evolution of a telescope's PSF, the best correlated reference images are usually the most contaminated by the planet, leading to self-subtraction and reducing the planet throughput.
    In this paper, we present an algorithm that directly optimizes the non-linear equation for planet signal to noise ratio (SNR). This new algorithm does not require us to reject adjacent reference images and optimally balances noise reduction with self-subtraction. We then show how this algorithm can be applied to multiple images simultaneously for a further reduction in correlated noise, directly maximizing the SNR of the final combined image.
    Finally, we demonstrate the technique on an illustrative sequence of HR8799 using the new Julia-based Signal to Noise Analysis Pipeline (SNAP). We show that SNR optimization can provide up to a $5\times$ improvement in contrast close to the star.
    Applicable to both new and archival data, this technique will allow for the detection of fainter, lower mass, and closer in companions, or achieve the same sensitivity with less telescope time.

\end{abstract}



\section{Introduction} \label{sec:intro}

Direct imaging of exoplanets offers astronomers a wealth of information -- from initial detection and astrometry, to detailed orbital and spectroscopic characterization. Due to the challenges of detecting faint companions close to their stars, direct imaging is currently limited in most cases to large, self luminous planets on wide orbits. A major goal in the field of direct imaging is thus to improve the planet-to-star contrast ratio to image fainter planets, closer to their stars.
Evidence from radial-velocity and direct imaging surveys suggest that the occurrence rate of giant planets peaks near $2-3 \;\mathrm{AU}$ and increases with lower mass \citep{fernandesHintsTurnoverSnow2019a,nielsenGeminiPlanetImager2020}. Improving contrast close to stars is therefore a clear path towards more direct detections.

Despite significant progress over the past decade, the dominant source of noise in direct imaging today remains quasi-static speckle noise. This noise represents aberrations in the point spread function (PSF) of the atmosphere, telescope, and instrument that are not corrected by the adaptive optics system, and have lifetimes on the order of seconds to tens of minutes. These quasi-static speckles are sufficiently correlated between images that simply averaging data over longer and longer integrations does not significantly improve the final contrast \citep{walkerShadesBlackSearching1998,maroisEffectsQuasiStaticAberrations2003}.
Since speckle fields can be filtered until only the component that appears planet-like remains, these residual, long living speckles appear identical to planets and can be much brighter.

For quasi-static speckles, most observing strategies depend on differential imaging. These are a range of techniques, most notably angular differential imaging \citep[ADI,][]{maroisAngularDifferentialImaging2006} and spectral differential imaging \citep[SDI,][]{walkerShadesBlackSearching1998,racineSpeckleNoiseDetection1999,maroisEfficientSpeckleNoise2000} that produce a geometric offset that varies from image to image between the stellar PSF and any planets. These images taken close in time and / or wavelength can then serve to build a model of the stellar PSF that ideally is representative of the speckle noise at each moment.

Most current algorithms based on LOCI \citep{lafreniereNewAlgorithmPointSpread2007} reduce differential imaging sequences by considering one image at a time---the ``target image''---and select a set of ``reference images'' from the other images of the sequence. Then, they find the linear combination of reference images that, when subtracted from the target image, create a ``processed image'' with minimial residual noise. The images are usually divided into small annular ``subtraction regions'' so that the model is specialized to the local noise distribution in the target image. Finally, they transform all of the processed images to align the signals of any planets and then stack them to produce the final output.

There are three challenges with this approach that must be given careful consideration: over-subtraction, overfitting, and self-subtraction. First, over-subtraction occurs when an algorithm fits the signal of the planet in the target image using some combination of speckle noise in the reference images and reduces planet throughput. This can be mitigated by running the algorithm on a separate ``optimization region'' paired with each subtraction region, and separating the two by a small buffer to prevent the linear least-squares solution from incorporating any planets into its model. When the subtraction and optimization regions are sufficiently close together, the speckle noise is typically correlated and the model created using the optimization region is effective at removing speckles in the subtraction region.

Second, when the sutbraction and optimization regions are separated, overfitting occurs when the model becomes too specialized to the optimization region and fails to generalize to the subtraction region, producing a model that does not remove speckle noise effectively. There are several ways to control overfitting, but the most common strategies all serve to reduce the dimensionality of the model. This can be done by only selecting the reference images that are the most correlated with the target image \citep[LOCI,][]{lafreniereNewAlgorithmPointSpread2007,maroisExoplanetImagingLOCI2010} or by linearly transforming the images into an orthogonal basis and only keeping the eigenimages with the highest variance \citep[KLIP/PCA,][]{soummerDetectionCharacterizationExoplanets2012}. The size of the optimization regions and the best number of reference images/eigenimages to include are usually correlated, with larger optimization regions being less specialized to the noise near the subtraction region, but allowing the use of more reference images before overfitting occurs.

Finally, and perhaps most importantly, planet self-subtraction occurs when the signal of a planet is suppressed by reference images contaminated with that same planet. Linear methods have a strong tendency towards selecting the most contaminated reference images in order to minimize the noise, since these are the images that are the most correlated with the target image. Traditionally, these reference images are rejected by applying a threshold to the local planet displacement between images (e.g. due to varying paralactic over time), the expected amount of flux contamination via forward modeling~\citep[TLOCI,][]{maroisExoplanetImagingLOCI2010}, or other similar criteria. This limits self-subtraction at the expense of increased noise, so this parameter (sometimes called the aggressiveness) and other hyper-parameters are varied to find the combation that maximizes the SNR of injected, forward modeled planets \citep[][and others]{lafreniereNewAlgorithmPointSpread2007,maroisExoplanetImagingLOCI2010,soummerDetectionCharacterizationExoplanets2012, meshkatOptimizedPrincipalComponent2014}. Some amount of self-subtraction is expected from these methods, so planet throughput must be calibrated by injecting forward modeled planets into the raw images.

Rejecting some of the most correlated references, followed by minimizing the noise, and then correcting the planet throughput might be sub-optimal. One improvement to this was considered in \citet{pueyoApplicationDampedLocally2012} in the Damped-LOCI algorithm. In addition to limiting spectral cross-talk when extracting spectra from SDI sequences, Damped-LOCI aims to improve planet SNR by simultaneously minimizing noise in the optimization region and maximizing the variance in the subtraction region, with an additional tunable hyper-parameter to balance the tradeoff. One possible issue with this approach is that by attempting to boost the variance in the subtraction region, speckle noise under the planet might be amplified.

To directly achieve an optimal planet signal to noise ratio, we present a new algorithm in this paper that re-frames the problem from minimizing the noise to directly maximizing the signal to noise ratio (SNR) of any point sources using the same forward modeling facilities implemented in most pipelines for throughput correction. This relatively straightforward approach allows us to use any reference image regardless of it's proximity to or flux contamination with the target image. This removes the need to reject the most correlated reference images, and removes the need to correct throughput loss due to self-subtraction. Finally, this formulation can be generalized to treat multiple target images simultaneously so that planet SNR in the overall stack of processed images is optimized, further suppressing correlated noise. This approach is a general algorithm for maximizing the signal to noise ratio when combining measurements with correlated noise, and a known variation in signal intensity.

We begin by describing the new SNR optimization algorithm. We show how it relates to linear methods like LOCI in a limiting case, and why it should be expected to meet or exceed the performance of linear methods even when hyper-parameters of those methods are tuned to maximize SNR. Next, we show how the algorithm can be generalized to optimize a grid of multiple target and reference images simultaneously and describe how this can further reduce correlated noise in some sequences. Finally, we analyze the performance of the SNR optimization algorithm on a sample ADI sequence.

\section{SNR Optimization}\label{sec:snr-opt}

\begin{figure}
    \centering
    \includegraphics[width=0.35\textwidth]{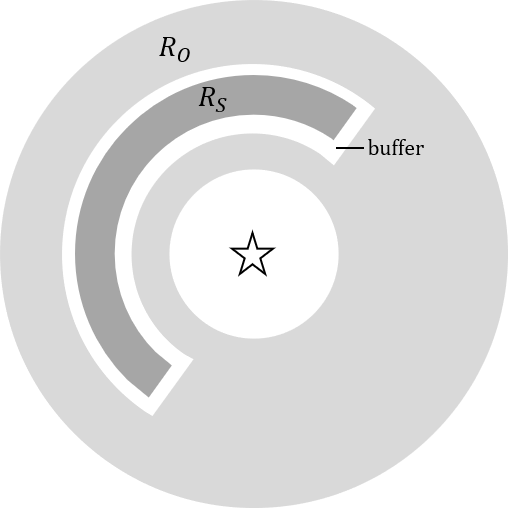}
    \caption{Examples of the region geometry used this paper. The annulus sector $R_S$ is the subtraction region and $R_O$ is the optimization region. The optimization region is chosen to be slightly thinner on the side closer to the star since the brighter speckles in that area could bias the model towards an overly aggressive solution. The white area between the two regions is a buffer to prevent any planets from leaking into the optimization region and causing over-subtraction.  The geometry of the regions can be chosen differently as long a sufficient buffer is preserved between them to prevent over-subtraction.}
    \label{fig:zone-geometry}
\end{figure}

To achieve an optimal planet signal to noise ratio, we propose directly optimizing an expression for SNR in each region of the sequence, instead of only minimizing the noise.
For a given location in the sequence, if we know the relative photometry of a planet in each input image by forward modeling, we can write a formula for the planet SNR for any linear combination of images.

First, we divide the images into pairs of optimization regions, $R_O$, and subtraction regions, $R_S$, just as in previous methods (Figure \ref{fig:zone-geometry}).
For each pair of regions, we consider a forward modeled planet centered in $R_S$ and noise evaluated in $R_O$.
We use the noise in the optimization region to find a solution that we then apply to the subtraction region. This makes the assumption that the noise in the two regions are correlated, but is necessary to avoid finding solutions that fit the signals of any planets using noise from the other images (called planet over-subtraction).
In previous works, authors have considered regions that partially overlap \citep[beginning with ][]{lafreniereNewAlgorithmPointSpread2007}; however, the methods presented here are aggressive enough that a buffer larger than approximately $0.4\lambda/D$ should be kept between the two regions to prevent over-subtraction.
Though we use the noise in the optimization region, we are free to model photometry for a planet located at the center of the subtraction region in the target image.

The signal of a linear combination of images is then simply the linear combination of the forward modeled relative photometries in each image:
\begin{equation}
    \mathrm{signal}=\mathbf{p}\cdot\mathbf{c},
\end{equation} where $\mathbf{p}$ is a vector of forward modeled relative planet photometry, and $\mathbf{c}$ is a vector of coefficients for the linear combination of images. The forward model should be calculated using  either satellite spots \citep{maroisAccurateAstrometryPhotometry2006,sivaramakrishnanAstrometryPhotometryCoronagraphs2006} on instruments where they are present, or an unocculted  and unsaturated PSF taken before or after the sequence.
This forward modeling is already implemented in most direct imaging reduction pipelines for the purpose of throughput correction \citep[e.g.][]{maroisExoplanetImagingLOCI2010, ruffioImprovingAssessingPlanet2017}.

By assuming that the noise is centered around zero, we can write the noise as the root-mean-square (RMS) of the pixel values in the same linear combination of images
\begin{equation}
    \mathrm{RMS} = \sqrt{\sum_{j}^M (\sum_i^N O_{ij} c_i)^2/M}
\end{equation}
where $i$ ranges from 1 to $N$, the number of images; $j$ ranges from 1 to $M$, the number of pixels; and $O$ is a matrix of $N$ images by $M$ pixels.
The RMS is chosen for simplicity and should be equivalent to the standard deviation if a radial profile is pre-subtracted, or the images are high-pass filtered. Note that this does not assume the underlying noise distribution is Gaussian, since we are not using the RMS to calculate confidence intervals. Minimizing the RMS of any reasonable distribution should minimize the noise in the images. It is possible that this metric could be tweaked to, for example, penalize spatially correlated noise, but this is outside the scope of this paper.

By combining these expressions, we arrive at an equation for the signal to noise ratio of the modeled planet for any given linear combination of images:

\begin{equation}\label{eq:snr}
    \mathrm{SNR} = \frac{\mathrm{signal}}{\mathrm{RMS}}= \frac{\sum_i^Np_i c_i}{\sqrt{\sum_{j}^M (\sum_i^N O_{ij} c_i)^2/M}}
\end{equation}

By noting that maximizing the SNR is equivalent to maximizing an expression proportional to the SNR$^2$, we can see that this optimization problem is a quadratically constrained quadratic problem:
\begin{eqnarray}
    \mathrm{argmax_c\;SNR}
    &=& \mathrm{argmax_c \; SNR^2}\\
    &=& \mathrm{argmax_c}\; \frac{
        \left(\sum_i^Np_i c_i\right)^2
    }{
        \sum_{j}^M (\sum_i^N O_{ij} c_i)^2
    }
    \label{eq:snr-squared}
\end{eqnarray}

The values of the coefficient vector $\mathbf{c}$ can then be chosen such that the planet SNR is maximized. Applying that vector of coefficients to the subtraction region produces the linear combination of images that gives close to the optimal planet SNR for that region, limited only by the degree of correlation between the chosen optimization and subtraction regions.

Since the SNR ratio is preserved if the $\mathbf{c}$ vector is scaled by a constant, the problem is under constrained. We therefore apply the additional constraint that the planet-throughput is kept at unity, or
\begin{equation}
    \mathbf{p}\cdot\mathbf{c}=1
    \label{eq:through-const}
\end{equation}
Then, if the forward modeling is done with a correctly scaled image of the same star, the resulting processed image is then naturally in units of relative contrast. This assumes that the planet forward model accurately reflects changes in, for example, atmospheric variations, or that these effects are small.

\subsection{Formulation as a constrained least-squares problem}

In order for this formulation to be useful, it must be possible to efficiently optimize the coefficients to find a global optimum.
With the application of the constraint in Equation (\ref{eq:through-const}), that is $\mathbf{c\cdot p}=1$, we can express this optimization problem as a constrained system of linear equations as follows:

\begin{eqnarray}
    \mathrm{argmax_c}\;\mathrm{SNR}
    &=& \mathrm{argmax_c}\; \frac{1}{\sqrt{\sum_{j}^M (\sum_i^N O_{ij} c_i)^2}} \\
    &=& \mathrm{argmin_c}\; \sqrt{\sum_{j}^M (\sum_i^N O_{ij} c_i)^2 }\\
    &=& \mathrm{argmin_c} \left|\mathbf{Oc}\right|\\
    &&  \mathrm{subject \; to \;} \mathbf{c\cdot p}=1
    \label{eq:dcp}
\end{eqnarray}


This finds the linear combination of reference images that minimizes the noise in the optimization zone, subject to the constraint that the throughput is preserved at unity. Without the constraint, the solution would have all $\mathbf{c}=0$ and the output image would be entirely empty. With this formulation, the photometry vector $\mathbf{p}$  decides which is the ``target'' image. The target image has its own coefficient, just like a reference.
Since we chose to model the planet photometry with a planet centered in the subtraction zone of the target image, the index of the target image should correspond to the maximum value of $\mathbf{p}$, and likely (though not necessarily) the highest value in $\mathbf{c}$ as well.

This formulation is a valid expression in the methodology of Disciplined Convex Programming \citep[DCP,][]{grantDisciplinedConvexProgramming2006}. Many programming languages have libraries that can solve such expressions efficiently to a global optimum.

In some cases, such as when using Reference Star Differential Imaging (RSDI) to ``star-hop'' between two stars, images with significant flux contamination and images with no contamination are intermixed. In this case, the constraints can be treated as sparse since only images with significant planet flux contribution must enter into the constraint term.

\subsection{Reduction to Linear Least Squares in limit of no self-subtraction}
To show how this method relates to previous algorithms, we show
that SNR optimization reduces to a linear problem if we neglect planet contamination in the reference images. This situation would occur if a target image was reduced using only reference images taken from another star (RSDI) or if all neighboring reference images were rejected (e.g. in LOCI based algorithms). Taking the target image to be $i=1$, then $p_1=1$, all other $p_i=0$, and $c_1=1$ due to the throughput constraint. Equation (\ref{eq:snr-squared}) then reduces to:
\begin{eqnarray}
    \mathrm{argmax_c}\;& \frac{1}{\sum_{j=1}^M \left(\sum_{i=1}^N O_{ij} c_i\right)^2}\\
    =\mathrm{argmin_c}&\sum_{j=1}^M \left( O_{1j}+\sum_{i=2}^N O_{ij} c_i\right)^2
\end{eqnarray}

Continuing now as in \cite{lafreniereNewAlgorithmPointSpread2007}, the minimum occurs when all of the partial derivatives with respect to $\mathbf{c}$ are zero:

\begin{eqnarray}
    \frac{\delta}{\delta c_k}\sum_{j=1}^M \left( O_{1j}+\sum_{i=2}^N O_{ij} c_i\right)^2 &=& 0,  \forall k\\
\end{eqnarray}
\begin{equation}
    \sum_{j=1}^M{O_{kj}O_{1j}} = \sum_{i=2}^N{-c_i}{\left(\sum_{j=1}^M{O_{ij}O_{kj}}\right)}, \;\; \forall k
\end{equation}

\begin{minipage}{\columnwidth}
    Which is the familiar system of linear equations of the form $\mathbf{Ax}=\mathbf{b}$ where\\
    \begin{equation}
        A_{ik}=\sum_{j=1}^M{O_{ij}O_{kj}},  \quad   x_k=-c_k, \ \ \mathrm{and} \ \ b_k=\sum_{j=1}^M{O_{kj}O_{1j}}.
    \end{equation}
\end{minipage}

This is exactly equation 7 of \cite{lafreniereNewAlgorithmPointSpread2007} apart from notational differences and the sign convention on $\mathbf{c}$.
We can therefore consider SNR optimization to be a generalization of LOCI for reference images that contain overlapping flux from the planet. Since this occurs in almost every ADI or SDI reduction, we can expect the resulting SNR to always be higher when using SNR optimization for the same set of references---ignoring secondary effects such as how well the solution generalizes from the optimization region to the subtraction region. Note that SNR optimization could be used directly on the subtraction regions analogously to some LOCI pipelines; however, in this paper we maintain a buffer to prevent oversubtraction.
Example target images reduced with both LOCI and SNR optimization are presented in Section~\ref{sec:demonstration}.

This also shows that we cannot further improve RSDI sequences since the two formulations would reduce to the same system of linear equations. If a sequence also contains field of view rotation or multiple wavelengths in addition to the reference images, this algorithm can still offer improved SNR by better handling the images that do have planet flux contamination.

\begin{figure*}
    \centering
    \includegraphics[width=0.7\textwidth]{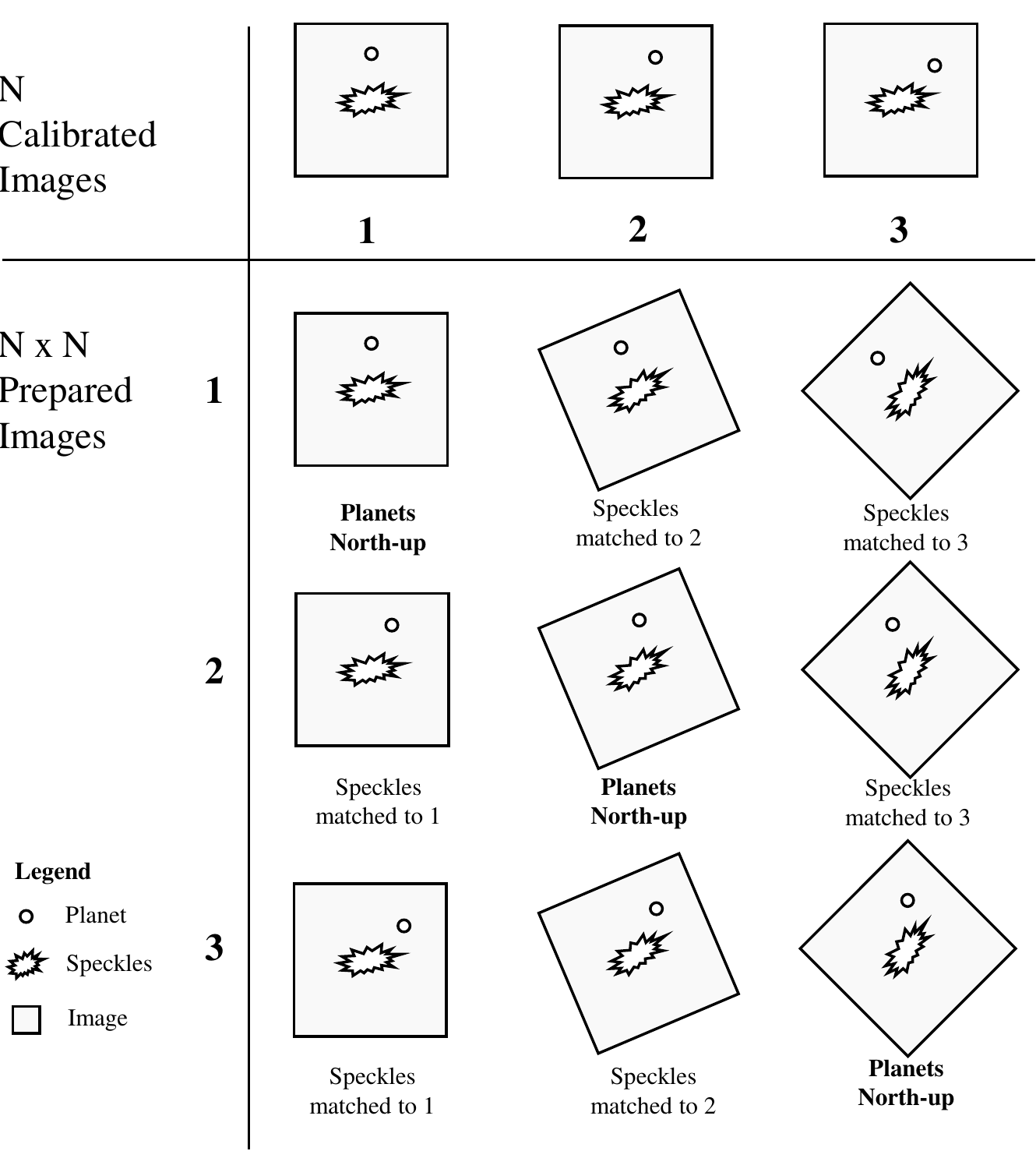}
    \caption{Preparing a grid of $N$ images for multi-target SNR optimization. In this example of a simplified 3-image ADI sequence, a distorted star shape represents the star's PSF and a small circle represents a planet directly to the North of the star.
        We begin by transforming each target such that any companion would appear in their physical, e.g. North-up, locations (the diagonal elements).
        Next, for each transformed target, we apply the same transformation to each corresponding reference (the off-diagonals). This aligns the speckles in the references with the speckles in that transformed target.
        We then follow the usual approach of dividing the field into annuli or annulus sectors for local optimization.
        An analogous matrix is prepared for the relative photometry using forward modeling for each region.
        In previous approaches, one would effectively reduce each column of this grid independently and then stack the results. By preparing this grid of transformed images as a single problem, the optimizer becomes aware of noise that is correlated between target images, and we achieve a higher point source SNR in the final stacked image.
    }\label{fig:whole-sequence-prep}
\end{figure*}

\section{Multi-Target Image Optimization}\label{sec:multi-opt}

In Section~\ref{sec:snr-opt}, we described the process of applying SNR optimization to a single target image with a set of reference images. The results of each image must then be rotated and/or scaled to re-align the signals of any planets and then stacked, just as in previous approaches. A limitation of this single-target optimization is the assumption that solutions giving the highest SNR for an individual target image also lead to the highest SNR in the stacked image.

This is a reasonable assumption to first order, but is often violated at close separations. For example, there are often bright, randomly fluctuating speckles in addition to a floor of dimmer, more consistent speckles. When considering each target individually, the SNR is maximized by suppressing the bright, fluctuating speckles, perhaps at the expense of the dimmer ones; however, in the final stack, it may be the dimmer persistent speckles that are correlated between images that contribute the most noise.

It is therefore worth considering ways to avoid this assumption. Since the SNR equations above do not distinguish between target images and reference images except by how the forward modeled photometry is calculated, it is possible to combine multiple target images into a single optimization problem. The output of the optimization algorithm then becomes a single, larger set of coefficients that describes both how each target image is combined with its matching references, and how those resulting images are combined into the final processed image.

There are three steps to preparing this problem. First, transform the target images to align any planet signals (usually by rotating them North-up). Then, transform copies of each reference image so that its speckles are aligned with each transformed target image. This is best visualized as a grid of reference images with the target images placed along the diagonal (see Figure \ref{fig:whole-sequence-prep}). Finally, generate subtraction and optimization regions that are in physical coordinates (e.g. North-up), rather than in the coordinates of the detector.

At this point, the algorithm can continue as usual, with a few adjustments to how the forward modeled photometry is calculated to account for those coordinate transformations. If all the target images are combined into a single problem, then there is no need to stack the results using a median or other statistic. Applying the coefficients to the subtraction region produces a single output image.

At first glance, this appears to increase the size of the model from $N$ images or eigenimages to $N^2$ images; however, the total number of coefficients to calculate is the same. Looking again at Figure \ref{fig:whole-sequence-prep}, standard linear methods or SNR optimization would calculate a coefficient for each image, on $N$-image column at a time, and the then stack the resulting processed images. If an arithmetic mean is used to stack the residuals, it's clear that the output consists of a linear combination of $N\times N$ images in either case.

By directly optimizing the SNR of that combination, we no longer assume that each target image is independent, and that the brightest noise, regardless of its temporal evolution, contributes the most to the final result. Secondly, we also use information in the optimization region to guide how much weight should be given to each column of the grid, rather than giving each target image an equal weight.
A comparison of the formance of single-target versus multi-target SNR optimization is presented in Section \ref{sec:demonstration}.

\subsection{Batching}\label{sec:batching}

Multi-target SNR optimization is useful for improving noise that is correlated between target images, and that is correlated between the subtraction and optimization region, but that does not dominate until multiple reduced images are stacked. On the other hand, a traditional stacking technique like a median is more effective at rejecting large outliers that are not spatially and temporally correlated since it operates inside the subtraction region.

These two techniques can be balanced by dividing the grid into groups of columns called batches.
Optimization can proceed one batch at a time to produce a smaller set of output images. The residuals can then be stacked analogously to per-target optimization, though without needing to first rotate or otherwise transform the results.
By grouping correlated targets together and then stacking the results, we retain most of the correlated noise rejection and the outlier rejection of a median.  We can then see that multi-target SNR optimization is a spectrum, ranging from considering only a single target at a time to an entire sequence.
For the same contrast, smaller batches have an additional benefit of reducing the computational cost of the reduction.

In principle, a correlation analysis could be used to group target images into batches, but a simple method is to group targets according to the time they were captured since images captured close in time are likely to be the most correlated.
The ideal size of a batch depends on both the sequence and instrument. It should be sufficiently large that target images outside the batch are no longer highly correlated, but of sufficient number that outliers can be rejected with a median.
This parameter can be optimized for each sequence or estimated by first reducing the sequence using single-target optimization and estimating the residual speckle lifetime.

Finally, the target batches do not need to be disjoint. It is reasonable to have, for example, a batch size of fifteen targets that advances five target images per batch. This slightly improves robustness against outliers at the expense of increased computation time.
An example of batched of multi-target SNR optimization is presented in Section \ref{sec:demonstration}.

\subsection{Controlling Overfitting}\label{sec:overfit}

In the context of differential imaging, overfitting occurs when a model is given too much flexibility and insufficient data. For example, this can occur if hundreds of reference images are used with a small optimization region to reduce one target image. The large set of references may be combined in such a way that random read noise, photon noise, or a transient speckle in the optimization region is minimized, but the solution does not effectively generalize to the noise in the subtraction region. This limits the performance of an algorithm but should not bias the signal of any planets as long as sufficient buffers are maintained.

Overfitting can be limited in several ways. First, one can restrict the set of reference images to only include those most correlated to the target \citep{maroisExoplanetImagingLOCI2010}. Second, one can perform an SVD \citep{maroisExoplanetImagingLOCI2010} or PCA analysis \citep{soummerDetectionCharacterizationExoplanets2012}, and similarly restrict the set of eigenimages. Third, one can increase the size of the optimization regions  until there is sufficient data to train the model (though if the optimization regions are too large, the model may not be sufficiently specialized to the local noise distribution). Another option is to introduce a regularization term to the model to penalize overly complex combinations of coefficients. Lastly, one average multiple reductions with variations in their parameters to reduce the impact of overfitting (sometimes referred to as ``bagging'').

The challenge of overfitting is exacerbated with multi-target SNR optimization. When many target images are combined into a single problem, the size of the reference set must grow quadratically. Unfortunately, SNR optimization is a non-linear process that does not admit a simple PCA solution. Such a solution would need to decompose the system of quadratic equations into non-linear ``SNR modes.'' Instead, we approach the problem of overfitting by combining four of the above techniques. First, we eliminate poorly correlated reference images from the grid, reducing the model size. Next, we add an L2 regularization parameter into the objective. Finally, we divide the sequence of target images into overlapping batches, and average their results. Each of these hyper-parameters can be optimized in turn to arrive at a balance that is ideal for any given sequence. This tuning process can be relatively efficient when using an optimizer that supports warm-starting from a previous solution.

When tuning hyper-parameters, it is best to use a stand-in such as a separate validation region, or set of prepared images with rotation reversed. By tuning the algorithm on this different but statistically similar data, we ensure that the hyper-parameters are not biased towards combinations that, by chance, reduce the flux of any planets and diminish overall throughput.

\begin{algorithm*}
    \caption{
        Multi-Target Image SNR Optimization for ADI.\label{algo:snropt}}
    \begin{algorithmic}
    \Procedure{SNROpt}{images, regions, $N_\mathrm{batch}$, $N_\mathrm{refs}$, $\alpha$}
        
        \For(\Comment{\parbox[t]{.448\linewidth}{Prepare images}}){i $\in 1:N_\mathrm{images}$}{

            $\theta_\mathrm{targ} \leftarrow \mathrm{PA}(\mathrm{images[i]})$
            \For{j $\in 1:N_\mathrm{images}$}{
                
                $\mathrm{prepared}[\mathrm{i,j}] \leftarrow \mathrm{rotate}(\mathrm{images[j]}, \theta_\mathrm{targ})$
            }
        }
        \For(\Comment{\parbox[t]{.448\linewidth}{Iterate through batches}}){i $\in 1:N_\mathrm{batch}:N_\mathrm{images}$}{ 

            \For(\Comment{\parbox[t]{.4\linewidth}{For each target in batch, select references}}){k $\in 1:N_\mathrm{batch}$}{ 

                selected[k] $\leftarrow$ $N_\mathrm{refs}$ most correlated references in column k of batch
            }
            \ForEach(\Comment{\parbox[t]{.4\linewidth}{Iterate through each pair of regions}}){$(R_S,R_O)\; \in \mathrm{regions}$}{

                $\mathbf p \leftarrow$ model photometry of selected in $R_s$ for planet centered in targets
                
                $\mathbf O \leftarrow R_O$ regions of selected images
                
                $\mathbf c \leftarrow  \mathrm{argmin} \left|\mathbf{Oc} + \alpha \sum_i^N{c_i^2} \right| \;\;\;  \mathrm{subject \; to \;} \mathbf{c\cdot p}=1$ \Comment{\parbox[t]{.4\linewidth}{Optimize SNR in optimization region}}
                
                $\mathbf S \leftarrow R_S$ regions of selected images

                $\mathrm{out_i}[R_S] \leftarrow \mathbf{Sc} $ \Comment{\parbox[t]{.4\linewidth}{Apply coefficients to subtraction region}}

            }
        }
        
        Measure contrast curves of for each $\mathrm{out_i}$ and take the weighted median
    \EndProcedure
   \end{algorithmic}

    \vspace{6pt}
    The hyper-parameters $N_\mathrm{batch}$, $N_\mathrm{refs}$, and $\alpha$ should be optimized against either a separate validation region $R_v$ or by reversing the angle $\theta_\mathrm{targ}$. This ensures values are not selected that, by chance, hurt the throughput of any planets in $R_S$.
\end{algorithm*}

\section{Implementation}


The full algorithm for multi-target image SNR optimization with regularization is presented succinctly in Algorithm 1.
This takes both the number of most correlated references to include and the regularization parameter $\alpha$ to limit overfitting. It also takes the number of images in a batch to control the trade off between outlier rejection ($N_\mathrm{batch}=1$) and reducing correlated noise  ($N_\mathrm{batch}=N_\mathrm{images}$). These hyper-parameters should be tuned for each sequence. The image preparation steps on lines 4 and 5 can be generalized for any type of differential imaging. We do not include hyper-parameter optimization steps for the geometry of the subtraction and optimization regions, as these are physically motivated, relatively insensitive, and already discussed in previous works \citep{lafreniereNewAlgorithmPointSpread2007, currieDirectDetectionOrbit2012}

To apply the algorithm, we built a new Signal to Noise Analysis Pipeline (SNAP). Implemented in Julia~\citep{bezansonJuliaFastDynamic2012}, it can process sequences with RDI, ADI, SDI from Keck NIRC2 (PI: K. Matthews), VLT-SPHERE~\citep{beuzitSPHEREExoplanetImager2019}, GPI~\citep{macintoshFirstLightGemini2014}, and LBT LMIRCam~\citep{skrutskieLargeBinocularTelescope2010}. Single-image and multi-target image SNR optimization is supported along with LOCI and TLOCI-style algorithms for the sake of comparison. The LOCI style algorithms include optimization steps for the number of reference images and rejection threshold, making it roughly comparable to other ``optimized'' implementations like \citet{meshkatOptimizedPrincipalComponent2014}.

To abstract RDI, ADI, and SDI as a single process, we used the library \code{CoordinateTransforms.jl} to store arbitrary transformations between speckle-aligned coordinates and planet-aligned coordinates. For multi-target image SNR optimization, we combined these transformation objects with \code{WarpedView}s from the \code{ImageTransformations.jl} package. This allows us to avoid storing duplicates of each reference image transformed to match each target. When a given pixel of a reference image is required, the transformation and interpolation are performed on the fly. This reduces the number of prepared reference images stored in memory from $N^2$ to simply $N$, and removes much of the memory overhead for multi-target SNR optimization.

Forward modeling is also necessary for the SNR optimization algorithm. To this end, the pipeline uses the transformation objects to compute a model for a planet in each subtraction region for each second of the sequence to account for the rotational smearing that takes place during ADI integrations.
These abstractions allow the pipeline to straightforwardly optimize the stacked SNR of a sequence containing multiple wavelengths and rotation angles without knowing the details of the transformations applied to the images.

For the SNR optimization itself, we implemented two modules. The first uses either Newton's method or BFGS \citep{fletcherPracticalMethodsOptimization1987,mogensenOptimMathematicalOptimization2018} to directly optimize the SNR using Equation (\ref{eq:snr-squared}) and its partial derivatives. This performs very well on smaller problems up to a few hundred reference images. On larger problem sizes, \code{Convex.jl} \citep{udellConvexOptimizationJulia2014} with the \code{COSMO.jl} \citep{cosmojl} backend is used with Equation (\ref{eq:dcp}). This outperforms the direct solution, and is particularly efficient on multi-target image SNR optimization problems which may have Hessians with very high condition numbers ($>10^7$).

\section{Demonstration}\label{sec:demonstration}

\begin{figure*}
    \centering
    \includegraphics[width=\textwidth]{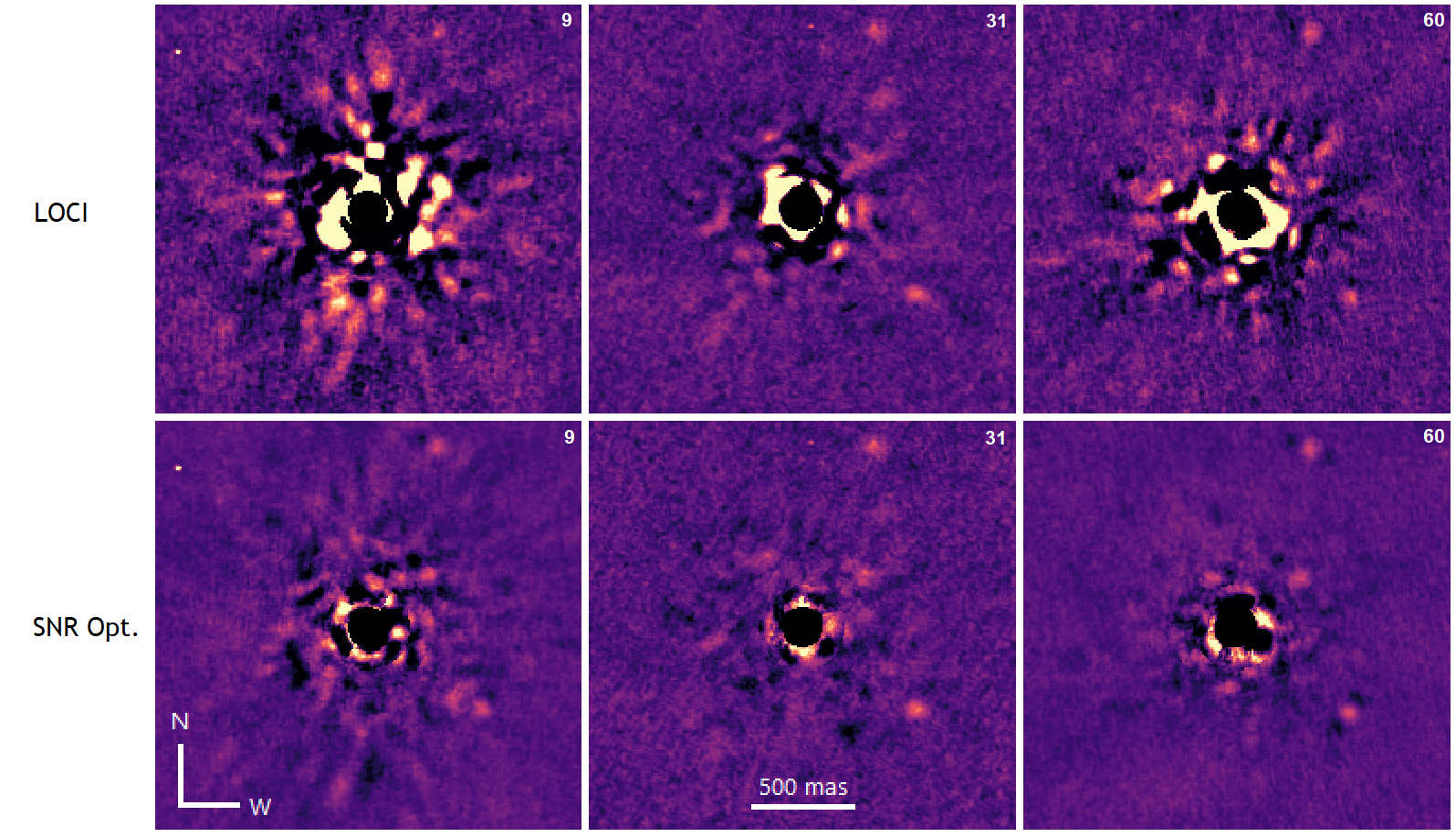}
    \caption{Three examples of target images reduced with both LOCI (top) and with SNR optimization (bottom). Each example was taken from a different part of the sequence with different FoV. rotation rates. In all three examples, we see that SNR optimization is very effective at suppressing speckles from the star. Images are displayed on the same color scale.}
    \label{fig:single-target-compare}
\end{figure*}

\begin{figure}
    \vskip12pt
    \centering
    \includegraphics[width=0.45\textwidth]{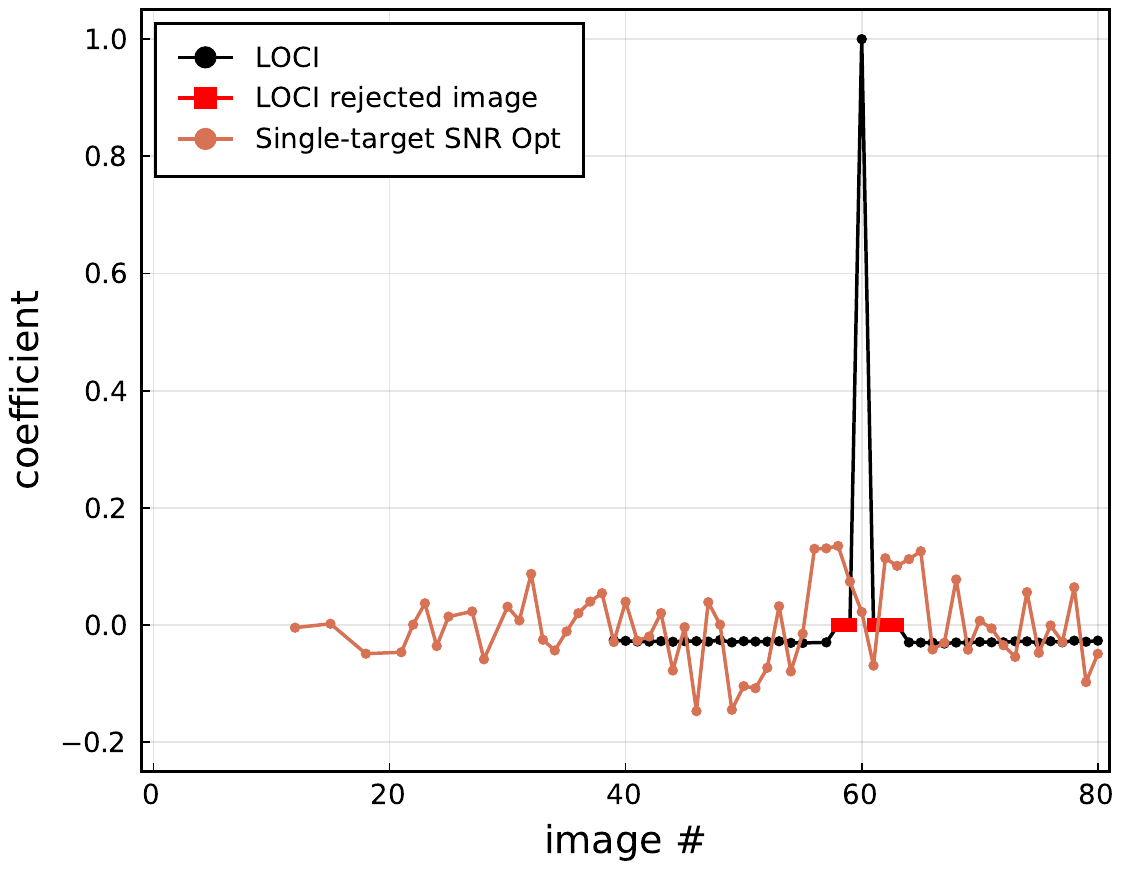}
    \caption{The coefficients chosen by LOCI and by SNR optimization for the images shown in the third column of Figure \ref{fig:single-target-compare} and a region close to planet HR8799e. The black line shows the coefficients chosen by the LOCI algorithm for the target image \# 60. The red squares indicate images that were rejected for the LOCI reduction due to insufficient displacement from the target. The orange line, on the other hand, shows the coefficients that give the highest SNR. Note that in SNR optimization, the target image is treated just like a reference image with a coefficient.}
    \label{fig:loci-vs-snr-coeff}
\end{figure}

\begin{figure*}
    \centering
    \includegraphics[width=\textwidth]{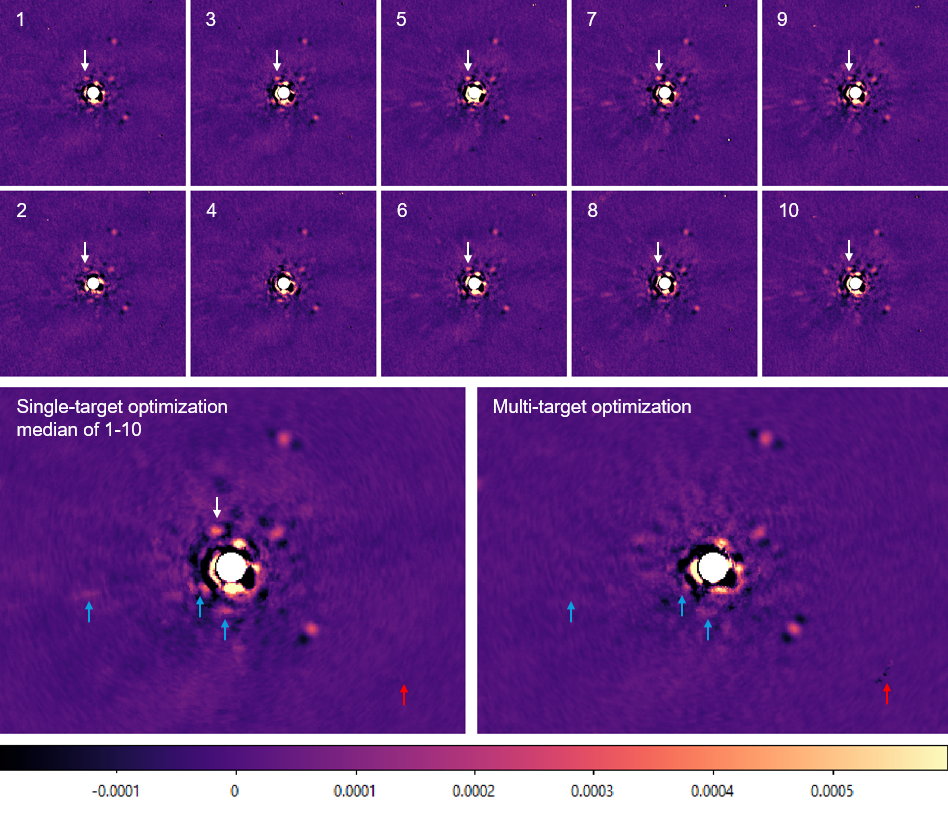}
    \caption{The benefits of multi-target SNR optimization on one batch of ten images.
        \textbf{Top:} Target images reduced individually with SNR optimization. A white arrow points to an example of noise that is correlated between nine out of the ten images.
        \textbf{Bottom left:} Median stack of the ten images at the top. The highly correlated residuals are not rejected by the median.
        \textbf{Bottom right:} The same images reduced simultaneously with multi-target image SNR optimization. By considering the ten targets simultaneously, we prevent the buildup of correlated noise between targets.
        The blue arrows highlight locations where noise that was correlated between the ten target images was suppressed by multi-target SNR optimization.
        The red arrow shows a counter example where multi-target SNR optimization fails to reject a bad pixel. Taking the median of multiple batches is an effective way to suppress these cosmetics.
        Images are in units of brightness relative to the star.
    }
    \label{fig:per-targ-vs-multi}
\end{figure*}

\begin{figure*}
    \centering
    \includegraphics[width=\textwidth]{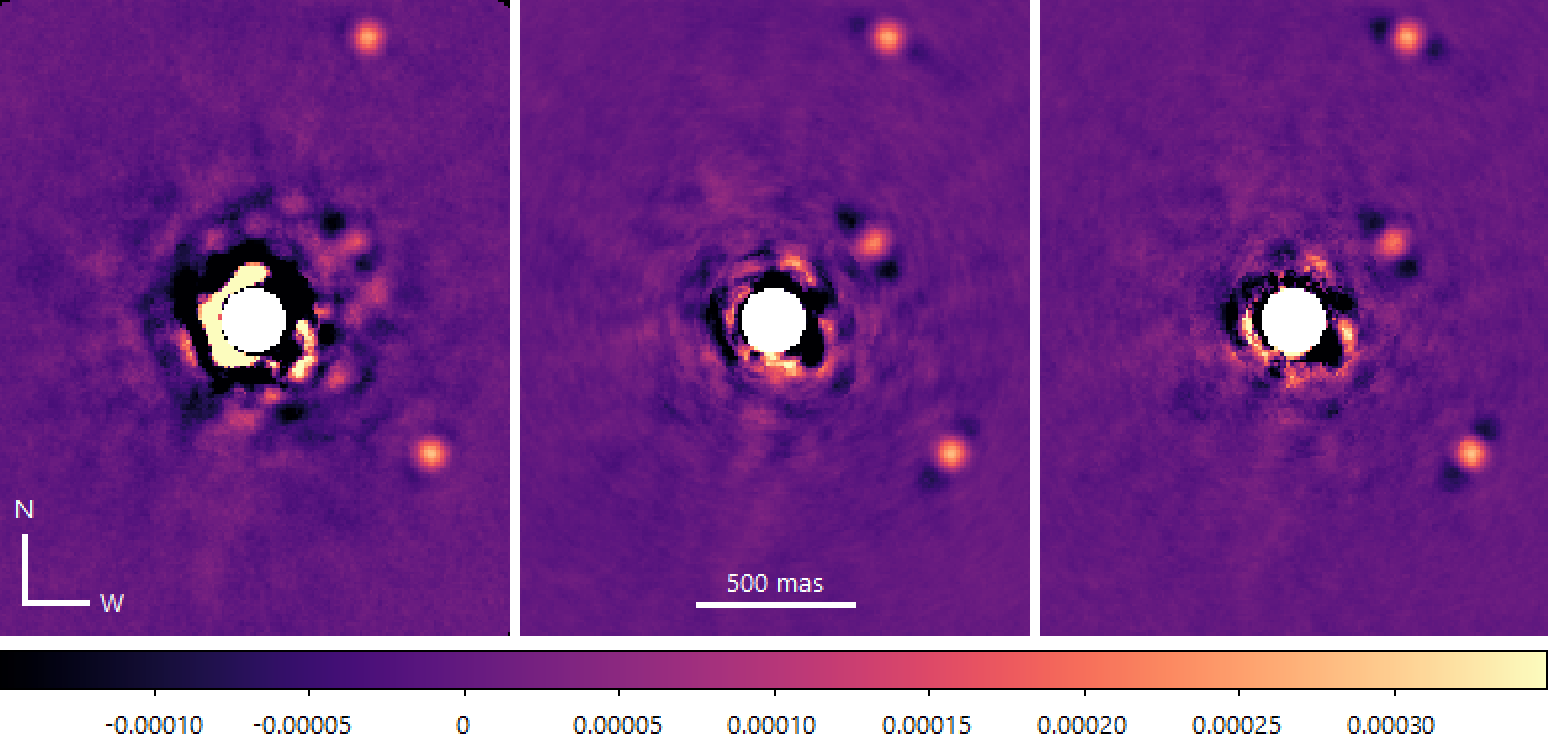}
    \caption{
        A non-coronagraphic L$^\prime$-band ADI sequence taken of HR8799 at Keck, reduced with LOCI (left), single-target image SNR optimization (middle), and multi-target image SNR optimization (right). While HR8799e is near the level of the noise in the LOCI reduction, the new algorithm recovers it robustly and even recovers usable data from behind the first Airy ring of the stellar PSF. The central $1.2 \; \lambda/D$ region marked in white is saturated.
        The outputs of the three algorithms are each combined in a contrast weighted median, which tends to favor images with the highest FoV rotation rate---hence the uncharacteristically faint dark wings in the LOCI reduction. Much of the central 500 mas in the left panel is outside the plotted color scale.}
    \label{fig:comparison-images}
\end{figure*}

\begin{figure}
    \centering
    \includegraphics[width=0.48\textwidth]{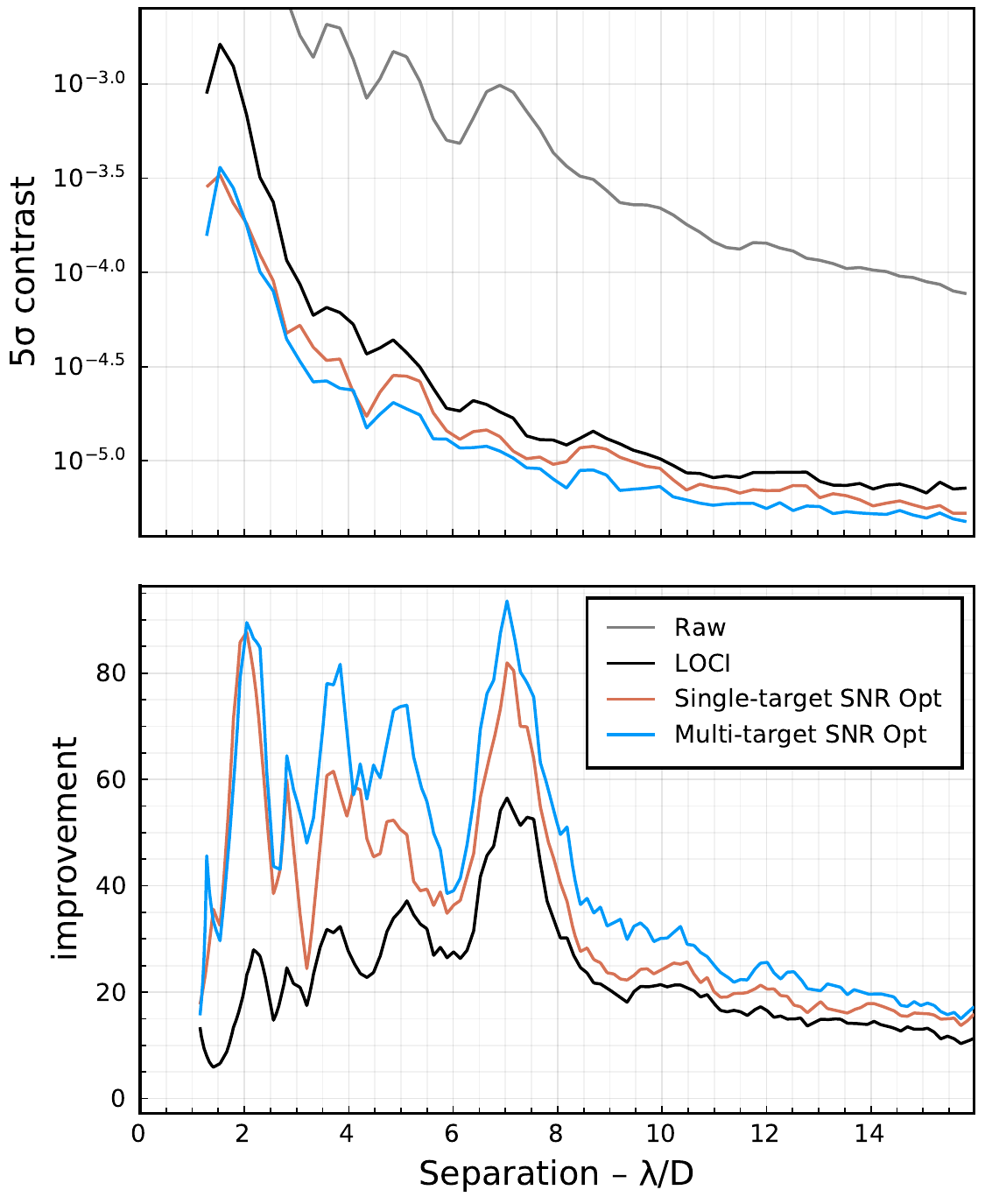}
    \caption{\textbf{Top:} planet-to-star contrast compared for the different reductions, and one raw image. \textbf{Bottom:} Contrast improvement compared to the raw image.
        Multi-target SNR optimization outperforms LOCI by three to five times at small separations from the star, and achieves a consistent 10-20\% improvement at wide separations.
    }
    \label{fig:comparison-contrast}
\end{figure}

\begin{figure*}
    \centering
    \includegraphics[width=\textwidth]{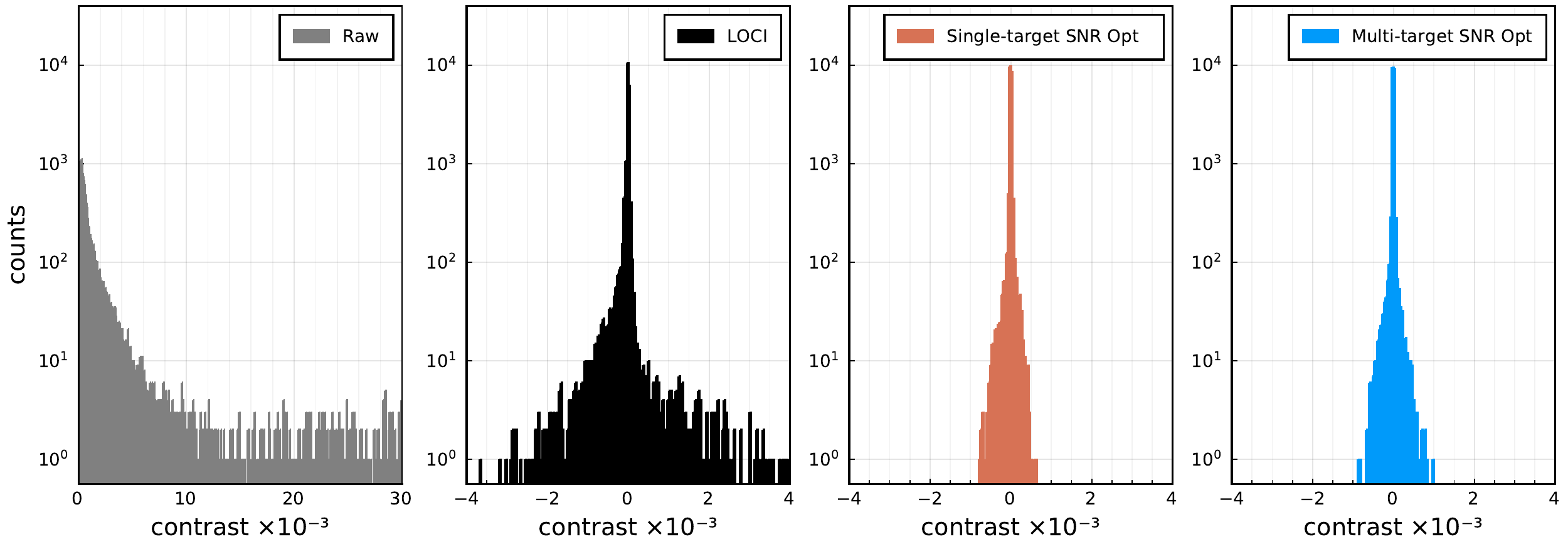}
    \caption{Histograms of residual noise in the inner $3\;\lambda/D$ region around the star for one raw image and the reduced images presented in Figure \ref{fig:comparison-images}. In addition to the lower overall noise, SNR optimization produces residuals with a more symmetric distribution on this sequence.
    }\label{fig:comparison-hist}
\end{figure*}

\begin{figure}
    \centering
    \includegraphics[width=0.48\textwidth]{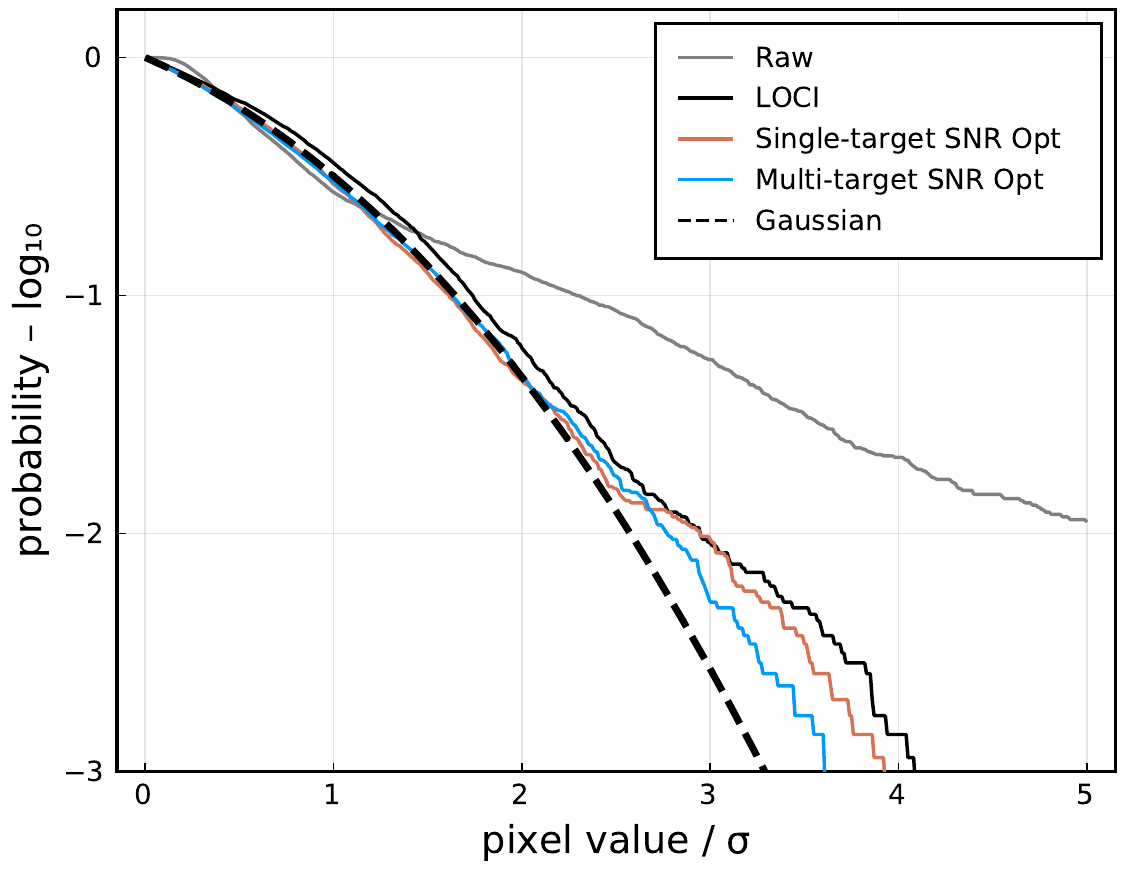}
    \caption{Comparison of the pixel distributions between the different algorithms and one raw image. The values shown are  1 - the cumulative distribution of pixel values taken from an annulus between 500 and 600 milliarseconds, normalized to one standard deviation and integrated symmetrically from zero. The algorithm presented in this paper produces residuals with a distribution that more closely resembles a Gaussian.}\label{fig:comparison-cdf}
\end{figure}

We evaluated the performance of the SNAP pipeline on an ADI sequence taken of HR8799 \citep{maroisDirectImagingMultiple2008a} with NIRC2 at Keck on 2020 November 17 (PI: Q. Konopacky). The sequence consists of 90 separate 40-second integrations taken at L$^\prime$ band while the system transited the meridian. We took the sequence in good seeing without a coronagraph, causing the central core of the star to saturate. We acquired separate unsaturated images at the beginning and end of the sequence.

We reduced the sequence three times using the same code, changing only whether the coefficients were selected by our baseline implementation of LOCI, by single-target image SNR optimization, or by multi-target image SNR optimization. This reduces the likelihood that any unrelated differences between implementations affect the results. For all three algorithms, hyper-parameters such as the number of reference images and the regularization were independently optimized to select their best values for this sequence. That said, general details such as the geometry of the optimization and subtraction regions were chosen to give the best performance using SNR optimization---it's possible that these, or other choices are not ideal values for LOCI style algorithms, and that comparable results could be achieved given sufficient effort. It is challenging to compare different algorithms across sequences and instruments in a completely fair way, so this should be considered only as an illustrative example for how SNR optimization performs.

For this sequence we used subtraction regions that are annulus sectors $1.4 \lambda/D$ thick and $180^{\circ}$ wide, and optimization regions that surround the subtraction regions with $2 \lambda/D$ on the outside, and $1.2 \lambda/D$ on the inner side. A buffer of $0.4 \lambda/D$ separates the two to prevent over-subtraction. See Figure \ref{fig:zone-geometry} for a schematic.

To tune the hyper-parameters of SNR optimization, we increased the number of included reference images and the L2 regularization parameter until the forward modeled SNR stopped improving. For LOCI, the rejection window and the number of included reference images were similarly adjusted.
We tuned the hyper-parameters using the backwards-rotated set of images to avoid unintentionally favoring solutions that bias planet photometry and risk losing planet throughput.
In both cases, the reference images were ranked against each target by their correlation to the subtraction region.
Finally, the LOCI images were throughput corrected using forward modeling in order to match the results of SNR optimization. This may lead it to look qualitatively worse than similar images in the literature where thoughput correction might be neglected but reflects the true ability of the algorithm to recover planets.

We begin by examining three individual target images, \# 9, 31, and 60, reduced with LOCI and with single-target image SNR optimization (Figure \ref{fig:single-target-compare}). The new algorithm significantly outperforms the baseline reduction and reveals the innermost known planet \citep{maroisImagesFourthPlanet2010}. In Figure \ref{fig:loci-vs-snr-coeff}, we compare the coefficients chosen by both algorithms for a subtraction region near planet HR8799e. For the sake of comparison, we adopt the convention that the target image has a coefficient held at 1 for the LOCI reduction. Reference images that lie within the LOCI rejection window are considered as having coefficients held at 0.

The difference in the chosen coefficients for the linear combination is striking. The images that were rejected by LOCI for having too much overlap are used to a significant extent by SNR optimization. Surprisingly, one of the directly adjacent reference images was subtracted from the target, despite significant planet overlap. This may indicate that SNR optimization is compensating for a poor quality image using its neighbors. This, and other surprising deviations from the coefficients chosen by LOCI shows that there is significant room for the new algorithm to outperform existing techniques.

Next, we compare ten target images reduced with single- and multi-target SNR optimization. Figure \ref{fig:per-targ-vs-multi} shows the ten images each reduced independently with SNR optimization. Many areas of the reduced images are highly correlated from image to image. The information necessary to remove these speckles must therefore be present in the reference images; however, it was not used. This could be because removing those speckles would have been a poor trade off for the SNR of a single reduced image. When stacked, these residual correlated speckles remain, limiting the SNR of the final image. When the ten target images are considered simultaneously, however, the objective function is sensitive to the final combined SNR, and therefore finds a better overall solution.
This shows that reducing correlated target images in batches prevents the build up of correlated residuals in the final stack.

Finally, we examine the full reduction of the sequence with LOCI and SNR optimization. We adopt a batch size of ten target images, and advance the window five targets at a time (i.e. producing 16 output images).
The processed images from LOCI and SNR optimization were stacked using a contrast-weighted median where the contrast was evaluated using a matching backwards-rotated reduction to prevent planet signals from biasing the weights. The final outputs of the two algorithms are shown in Figure \ref{fig:comparison-images}.
The images are cropped to the inner region of the system to highlight the improvement close to the star.

Planets b, c and d \citep{maroisDirectImagingMultiple2008a} are recovered in both reductions, while planet e \citep{maroisImagesFourthPlanet2010} is only recovered robustly from this sequence using the new algorithm.
Though the performance of both types of algorithm are comparable at wider separations, regions closer to the star are dramatically improved.
We hypothesize that the improved performance is due to its more optimal treatment of the reference images that overlap with the target image, and its use of multi-target SNR optimization to limit the build up of correlated noise.

The shapes of the planet PSFs and negative side-lobes are quite different between the two reductions, indicating that SNR optimization chooses very different coefficients. Unlike LOCI, SNR optimization is free to add adjacent reference images, or even improve the planet signal using the first Airy ring of the planet from a reference image. The net effect at small separations from the star is typically a broader planet PSF than LOCI.

Figure \ref{fig:comparison-contrast} shows the contrast improvement of each algorithm compared to a raw image.
We find that at wider separations, the contrast was improved by roughly 20\%, while at smaller separations, the contrast was improved by nearly 5 times.

Figure \ref{fig:comparison-hist} shows histograms of the pixel values of a region close to the star for the three fully reduced images. The spread of pixel values is much narrower in the SNR optimization reductions, and is also more symmetrical with fewer outliers. Figure \ref{fig:comparison-cdf} shows how, in addition to producing images with lower residual noise, SNR optimization can also produce residuals whose distribution more closely resembles a Gaussian. This reduces the effects of non-Gaussian noise on detection, astrometry, and photometry confidence intervals.

\section{Discussion}\label{sec:discussion}

Now, we discuss a few considerations when applying this technique and directions that warrant further study.

First, there may be sequences where LOCI outperforms single-target image SNR optimization because a more aggressive subtraction, while sub-optimal for each processed image, improves the final stacked image. The core of the SNR optimization algorithm is parameter-free, so it is not possible to tune the aggressiveness to replicate this effect. Instead, multi-target SNR optimization should be used so that an optimal solution for the final stacked image is found.
Compared to using LOCI with a tuned aggressiveness, this does rely more heavily on the assumption that the subtraction and optimization regions are well correlated.

Second, this technique does not permit a straightforward application of PCA methods which are useful for rejecting noise on some sequences. A potential future direction would be to transform the images into an approximately  orthogonal basis for the SNR function. In the meantime, an archive of images collected from reference stars or images where the planet PSF does not overlap with the target image could still be used to generate an orthogonal basis using an algorithm like KLIP. These eigenimages could then be used in combination with the overlapping reference images to combine the strengths of both algorithms. The use of a regularization parameter as described in Section \ref{sec:overfit} should achieve a similar result by favoring solutions with lower complexity.

Third, this method may distort a planet's PSF, shifting its apparent centroid and affecting its photometry. Just as with previous techniques, astrometry and photometry can be extracted robustly using forward modeling \citep[described in][]{maroisExoplanetImagingLOCI2010, galicherAstrometryPhotometryHigh2011}. Ultimately, it could be combined with a full forward-model matched filter~\cite{ruffioImprovingAssessingPlanet2017} to further improve the contrast. In general many additional steps and constraints have been shown to improve the performance of linear methods, and will likely be equally applicable to SNR optimization.

Finally, SNR optimization has a higher computational complexity than solving a system of linear equations. When combined with multi-target SNR optimization on a large sequence, the computational cost becomes significant. Our Julia-based SNAP implementation is able to reduce a single wavelength ADI sequence in roughly two hours on a 2015 Mac Pro workstation, or in just a few minutes when restricting the space explored by hyper-parameter tuning. Large SDI sequences can push the computational requirements higher such that a compute cluster becomes more appropriate.

\section{Conclusion}\label{sec:conclusion}

We have presented a new algorithm for processing high contrast images based the direct optimization of the non-linear SNR function. We showed that this approach improves on previous techniques that find linear least squares solutions that minimize the noise, even if the parameters like aggressiveness are optimized. The new algorithm no longer requires us to reject reference images taken too close in time or wavelength. We also showed how this new formulation allows us to reduce multiple target images simultaneously.
This allows the optimizer to work across the entire sequence, taking into account the temporal coherence of noise in addition to its spatial properties, and delivering optimized full-stack SNR.
Finally, we presented our implementation in SNAP, the Signal to Noise Analysis Pipeline, and demonstrated a significant improvement to contrast in the challenging region close to the star.
We expect this algorithm will improve the contrast of ground based facility-class instruments and future flagship space missions. This will enable the detection of fainter, lower mass, and closer in companions, or achieve the same sensitivity with less telescope time.

\section{Acknowledgments}

We thank Q. Konopacky for allowing us to participate in the Keck NIRC2 HR8799 L$^\prime$ campaign and use the data in this paper. We also thank the anonymous referee whose suggestions improved this paper. We acknowledge the support of the Natural Sciences and Engineering Research Council of Canada (NSERC) and New Technologies for Canadian Observatories (NTCO) program.
The data presented herein were obtained at the W. M. Keck Observatory, which is operated as a scientific partnership among the California Institute of Technology, the University of California and the National Aeronautics and Space Administration. The Observatory was made possible by the generous financial support of the W. M. Keck Foundation.
This research was enabled in part by support provided by WestGrid, Compute Ontario and Compute Canada.
The authors wish to recognize and acknowledge the very significant cultural role and reverence that the summit of Maunakea has always had within the indigenous Hawaiian community.  We are most fortunate to have the opportunity to conduct observations from this mountain.


\bibliography{SNROpt}{}
\bibliographystyle{aasjournal}

\clearpage



\end{document}